\documentclass{article}

\usepackage{latexsym}
\usepackage{amsmath}
\usepackage{amsthm}
\usepackage{amsfonts}

\newtheorem{theorem}{Theorem}

\def\bbs+{|+\rangle}

\def\bs+{$\bbs+$}

\begin{document}
\begin{center}
{\Large Analyzing the effectiveness of the quantum repeater}\\
\end{center}
\begin{center}
Kenichiro Furuta, Hirofumi Muratani, Taichi Isogai and Tomoko Yonemura\\
\end{center}
\begin{center}
Corporate Research \& Development Center, Toshiba Corporation,\\
1, Komukai-Toshiba-cho, Saiwai-ku, Kawasaki, 212-8582, JAPAN\\
E-mail: {\ttfamily \{furuta,muratani,isogai,yonemura\}@isl.rdc.toshiba.co.jp}
\end{center}

{\bf Abstracts}\hspace{0.5cm}The communication distance of QKD is 
limited by exponential attenuation of photons propagating through optical fibers. 
However, it has been shown that introducing a quantum repeater can improve the order of the attenuation 
and is useful for extending the communication distance of QKD when the repeater noise 
is ignored. 
In this paper, we analyze the effectiveness of the quantum repeater 
when taking the repeater noise into consideration. 
We analyze the effectiveness also from the viewpoint of the security and show that 
QKD is secure even if a quantum repeater is used.

\section{Introduction}
%\subsection{Importance of quantum key distribution}
%\paragraph{The security and importance of BB84}
There is an everlasting threat that 
a current practical cryptographic scheme whose security is based on computational assumptions
will become insecure due to a future improvement of computers. 
Therefore, {\it quantum key distribution} (QKD), e.g. BB84\cite{BB84} and B92, 
has been attracting considerable attention 
because its security is based only on quantum principles and it is unconditionally secure.

%\subsection{Obstacles for quantum key distribution}
%\paragraph{The importance of the single photon generator over the security}
Due to exponential attenuation of photons in the channel, 
the naive QKD is valid only in the range of short distance.
It is important to extend the communication distance from a practical viewpoint.
So far, three approaches have been proposed to extend the range of QKD:

1.Protocol modification for multiple photon emission:
Protocol modifications\cite{sarg} which make the scheme robust 
against the photon number splitting(PNS) attack were proposed. 
However, modified protocols can extend the range of QKD to only a few times that of the original. 
Due to this limit, further extensions require introduction of other improvements.

2.Coherent states:
Some protocol using coherent states\cite{y00} were proposed. 
It was demonstrated that they are more resistant to noises than the single photon protocols  
and can achieve high bit rate even in long distance. 
However, its security has been discussed enthusiastically\cite{againsty00p3,yuenfory00}. 
In this paper, we do not consider this.

3.Quantum repeater protocol:
In order to reduce noises on quantum state transferred through the optical fiber, 
quantum teleportation is used to send the quantum state by using 
an EPR pair generated by a quantum repeater protocol. 
%as in Fig.\ref{fig:repeater3}. 
We call such a scheme {\it QKD with quantum repeater}. 
The quantum repeater recursively applies 
{\it entanglement swapping}(ES) and {\it entanglement purification protocol}(EPP) 
to short-length-EPR pairs\cite{repeater}. 
%as in Fig.\ref{fig:repeater2}. 
It was demonstrated in \cite{repeater} that a quantum repeater protocol 
can generate the long-length-EPR pair for the quantum teleportation with high fidelity. 

At a glance, the quantum repeater seems to be the most promising 
approach of the three approaches. 
However, the discussion in \cite{repeater} seems to assume 
that noises in quantum memory on checkpoints can be negligible. 
Therefore, we reexamine the practical possibility of the quantum repeater 
by evaluating the order of the bit rate with considering the repeater noise. 
Although the repeater noise is noticed in \cite{concrete}, 
the evaluation is done without the repeater noise.
We also examine the security of QKD with quantum repeater.

In Section \ref{sec2}, we review the quantum repeater protocol in \cite{repeater}.
In section \ref{sec3}, we categorize noises that occur in the protocol.
In Section \ref{sec4}, we evaluate the bit rate without considering the repeater noise.
In Section \ref{sec5}, we evaluate the bit rate with considering the repeater noise. 
In Section \ref{sec6}, we prove the security of QKD with quantum repeater.
In Section \ref{sec7}, we provide a summary of this paper.

\section{Quantum repeater protocol}\label{sec2}
The quantum repeater is a scheme which extends the length of an EPR pair
with high fidelity.
In this section, we review the quantum repeater protocol in \cite{repeater}.

\subsection{Abstract specification}
We explain an abstract specification of the quantum repeater protocol.
The protocol recursively applies ES and EPP 
to short-length-EPR pairs and 
finally generates a long-length-EPR pair with high fidelity. 

Let $L$ be the number of EPR pairs which are linked in a single ES execution, $N$ be 
the number of checkpoints and $n$ be the depth of the recursive executions.
These satisfy a relation, $N=L^{n}$.
In the channel between the sender $A$ and the receiver $B$, 
$N-1$ checkpoints, denoted $C_{1},C_{2},\cdots,C_{N-1}$, are settled.
For convenience, $A$ and $B$ are denoted $C_{0}$ and $C_{N}$, respectively.
The distance between $A$ and $B$ is denoted as $D$ and the distance between
two adjacent checkpoints is denoted as $d$. That is, $D=Nd$.

Then, quantum repeater protocol can be written as follows.
\begin{itemize}
\item Initialization: At each checkpoint $C_{i}$, 
$i=0,\cdots,$ $N-1$, EPR pairs are generated 
and one photon of each pair is sent to the next checkpoint $C_{i+1}$.
\item FOR{$x=1$ to $n$}
\begin{itemize}
\item ES: Execute ES in each of the checkpoints 
$C_{kL^{x-1}}$,$(k=1,2,\cdots,N/L^{x-1})$ 
except $C_{L^{x}},$ $C_{2L^{x}}$, $\cdots,C_{N-L^{x}}$.
Then, the EPR pairs of length $L^{x}$ can be obtained. 
\item EPP: Execute EPP for EPR pairs in each of the checkpoints\\ 
$C_{L^{x}},C_{2L^{x}},\cdots,C_{N-L^{x}}$.
\end{itemize}
\item Then, the EPR pairs of length $L^{x}$ with high fidelity can be obtained.
\end{itemize}
After completing the above protocol, an EPR pair with high fidelity shared between $C_{0}$ and $C_{N}$ is obtained.

\subsection{Entanglement swapping}
In ES, partners of two EPR pairs are swapped.
Here, we provide an explicit realizations of ES based on local measurement. 
ES can also be realized based on Bell measurement.

First, Controlled NOT gate (CNOT) is applied to photons 2 and 3 of 
$|\phi^{+}\rangle_{1,2}\otimes|\phi^{+}\rangle_{3,4}$.
Next, WH(Walsh-Hadamard) transformation is applied and produces
$\frac{1}{2}(|0\rangle_{2}|0\rangle_{3}\otimes|\phi^{+}\rangle_{1,4}
+|0\rangle_{2}|1\rangle_{3}\otimes|\psi^{+}\rangle_{1,4}
+|1\rangle_{2}|0\rangle_{3}\otimes|\phi^{-}\rangle_{1,4}
+|1\rangle_{2}|1\rangle_{3}\otimes|\psi^{-}\rangle_{1,4}).$
Then, one of four computational bases of photons 2 and 3 is measured 
and this measurement maps the state of two photons 1 and 4 into a Bell state.
Here, the observed basis of photons 2 and 3 indexes 
the projected Bell state of photons 1 and 4.
In the next step, the projected Bell state of photons 1 and 4 is transformed 
into $|\phi^{+}\rangle$.
For this purpose, the measurement results of photons 2 and 3 are 
sent from a checkpoint having photons 2 and 3 to 
ckechpoints having photons 1 and 4 with the classical communication. 

\subsection{Entanglement purification}
EPP pulls out an EPR pair of high fidelity from multiple EPR pairs of low fidelity.
We consider an EPP which is also considered in \cite{repeater}.

The validity of EPP requires that the fidelity of EPR pairs before the purification
should be in a certain range. It is demonstrated as follows.
Let $F$ and $F^{'}$ be the fidelity of the EPR pairs 
before EPP and the fidelity of purified EPR pair, respectively.
In the case that EPP generates the purified pair from two EPR pairs,
$F'$ can be expressed in terms of $F$ as follows\cite{purification}:
\begin{eqnarray}
\label{eq:afterpuri}
F'=\Phi / \Lambda,\hspace{0.1cm}where \bar{F}=(1-F)/3, 
\Phi=F^{2}+\bar{F}^{2} and \Lambda=F^{2}+2F\bar{F}+5\bar{F}^{2}.
\end{eqnarray}
In order that $F^{'} \geq F$ in Eq.(\ref{eq:afterpuri}), $F$ should be in the range of $1/2 \leq F \leq 1$.

\section{Noises}\label{sec3}
Here, we categorize possible noises which occur during an execution of the protocol.
\subsection{Noises during the protocol execution}
Several types of noises can occur during the execution of the quantum repeater protocol.
We classify them by their causes. 
The measurement noise is noises which occur during a measurement of a quantum state.
The one-qubit operation noise is noises which occur during a one-qubit operation in the protocol.
The two-qubit operation noise is noises which occur during a two-qubit operation in the protocol.
The channel noise is noises of a quantum state transferring through the channel.
The repeater noise is noises of a quantum state in the repeater devices even in the absence of operation.
Here, the one-qubit operation and the two-qubit operation mean a unitary operation on one qubit 
and a unitary operation on two qubits, respectively.
We consider that quantum noises which occur during the classical communication in the execution of 
EPP or quantum teleportation is an example of the repeater noise. 

The first three noises were modeled and analyzed in \cite{repeater}. 
For the channel noise and the repeater noise, we evaluate with the order, 
such as exponential or polynomial.
We review the models and analyses of the first three noises in the next subsection.

\subsection{Conventional models and analysis}
We show models of such noises and the modification of Eq.(\ref{eq:afterpuri}) caused by these noises.
Let $\rho$ be a density matrix before operations.
First, the one-qubit operation noise is modeled as 
$\rho \rightarrow O_{1} \rho = p_{1} O_{1}^{ideal} \rho +\frac{1-p_{1}}{2} tr_{1} \rho \otimes I_{1}$, 
where $O_{1}$ and $O_{1}^{ideal}$ are one-qubit operations with and without the noise, respectively,
and $I_{1}$ is the identity operator and
$p_{1}$ is the probability with which the operations are performed without noise.
The two-qubit operation noise is modeled as
$\rho \rightarrow O_{12} \rho = p_{2} O_{12}^{ideal} \rho+\frac{1-p_{2}}{4} tr_{12} \rho \otimes I_{12}$, 
where $O_{12}$ and $O_{12}^{ideal}$ are two-qubit operations with and without the noise, respectively,
and $I_{12}$ is the identity operator and
$p_{2}$ is the probability with which the operations are performed without noise.
The measurement noise is modeled as
$P_{0}^{\eta}=\eta|0\rangle \langle0| +(1-\eta)|1\rangle \langle1|,
P_{1}^{\eta}=\eta|1\rangle \langle1| +(1-\eta)|0\rangle \langle0|$, 
where $\eta$ is the probability with which the measurements are performed correctly 
and $P_{0}^{\eta}$ and $P_{1}^{\eta}$ are POVM $|0\rangle\langle0|$ 
and $|1\rangle\langle1|$, respectively, with error probability $\eta$. 

Based on the above noise models, the fidelity, $F_{L}$, 
after linking $L$ EPR pairs by ES executions is expressed as 
$F_{L}=\frac{1}{4}+\frac{3}{4} \left( \frac{p_{1}^{2}p_{2}
(4\eta^{2}-1)}{3} \right)^{L-1} \left( \frac{4F-1}{3} \right)^{L}.$
Similarly, based on the above noise models,
the change of the fidelity by EPP is expressed as follows:
\begin{eqnarray}
\label{eq:noiseafterpuri}
F'=
\{ \Theta\Phi +2\eta \bar{\eta}\Xi+\Pi \} /
\{ \Theta\Lambda+4(2\eta\bar{\eta}\Xi+\Pi) \},
\end{eqnarray}
where $\bar{\eta}=1-\eta$, $\Theta=\eta^{2}+\bar{\eta}^{2}$,  
$\Xi=F\bar{F}+\bar{F}^{2}$ and $\Pi=\frac{1-p_{2}^{2}}{8p_{2}^{2}}$.
$F$ and $F^{'}$ have three intersections. 
Let two intersections except 0.25 be $F_{min} and F_{max}$, 
where $F_{min}<F_{max}$.
Then, in order that $F^{'} \geq F$ in Eq.(\ref{eq:noiseafterpuri}), 
$F$ should be in the range of $F_{min} \leq F \leq F_{max}$.
The range of $F$ where the quantum repeater is valid become narrow 
as noises become large.

\section{Bit rate in absence of repeater noise}\label{sec4}
In \cite{repeater}, it was demonstrated that the required amount of resources 
of the quantum repeater increases as a polynomial function of 
the distance between $A$ and $B$, $D$.
This leads to the conclusion that the bit rate of the quantum repeater 
decreases as an inverse of a polynomial function of the distance.
This result was derived under the condition that 
only the channel noise, the measurement noise, 
the one-qubit operation noise and the two-qubit operation noise are considered.
\begin{theorem}\label{th1}
Consider QKD with quantum repeater.
If only the channel noise, the measurement noise, 
the one-qubit operation noise and the two-qubit operation noise are considered,
there exists a polynomial function $p(\cdot)$ such that 
the bit rate of the QKD decreases as $\Omega(p(D)^{-1})$.
\end{theorem}
Here, $g(n)=\Omega(f(n))$ means 
$\exists c>0 \, \exists N \in \mathbb{N}  \, 
\forall n>N \, g(n) \geq cf(n)$.

%In the following, $g(n)=O(f(n))$ means 
%$\exists c>0 \, \exists N  \, 
%\forall n >N \, g(n) \leq cf(n)$, where $c,N$ is a constant.

\begin{proof}
The bit rate is estimated by considering both the merit of quantum repeater, keeping 
high fidelity, and the demerit, increase of resource.

Let $M$ be the number of EPR pairs consumed by a single execution of EPP. 
Then, the number of EPR pairs, $R$, in the whole execution of the quantum repeater, 
is $R=(LM)^{n}=L^{n}M^{n}=NM^{n}=N(L^{\log_{L}M})^{n}=N^{\log_{L}M+1}$, 
where $N$ is proportional to the communication distance $D$, and $L$ and $M$ do not depend on $D$.
Thus, $R$ is a polynomial function of $D$.

In this scheme, when not considering the repeater noise, the fidelity of the EPR pair generated 
by the quantum repeater protocol stays constant even if $D$ increases.
So, the bit rate of the QKD decreases as $\Omega(p(D)^{-1})$.
\end{proof}

%The rate of photons which can convey information via optical fiber decreases exponentially.
%When considering only the channel noise, the measurement noise, the one-qubit operation noise, the two-qubit operation noise,
%and not considering the repeater noise, the bit rate of the quantum key distribution with quantum repeater decreases as the inverse of O(polynomial function) of the communication distance
%in contrast to the bit rate without quantum repeater decreases exponentially.
%Thus, the quantum repeater can be said to be effective considering only with these noises.
In contrast to exponential damping in the absence of the quantum repeater,
the bit rate of QKD in the presence of the quantum repeater decreases as an inverse of a polynomial function 
of the communication distance. Although Theorem \ref{th1} does not indicate 
whether the exact value of the bit rate is really improved by the quantum repeater,
it can be expected to be effective for sufficiently large $D$.

\section{Bit rate in presence of repeater noise}\label{sec5}
We next consider the case the repeater noise is taken into account.
In ES and EPP in the quantum repeater protocol,
classical communications between repeater devices are needed.
In addition, quantum teleportation sends classical information from $A$ to $B$.
During these classical communications, the quantum states in the repeater devices
lose their fidelity.
We assume the repeater noise as follows:
%\begin{assumption}[Exponential damping assumption]
The fidelity of a quantum state in a repeater device decreases exponentially
with respect to the time length of a classical communication.
%\end{assumption}
We call the assumption of this model {\it the exponential damping assumption}.
\begin{theorem}\label{th2}
Under the exponential damping assumption, 
the bit rate of QKD with quantum repeater decreases asymptotically exponentially 
with respect to the distance $D$. 
\end{theorem}
\begin{proof}
The length of EPR pairs increases as the quantum repeater protocol proceeds.
As far as the length is small, the fidelity can be recovered by EPP.
However, if the length exceeds a threshold, $D_{th}$, then the fidelity becomes 
lower than $F_{min}$ and EPP can't recover the fidelity any more. 
The reason is that EPR pairs have to stay in quantum memory on ckeckpoints 
during classical communication and the time of classical communication becomes large 
as the length of EPR pairs become large. Thus, time of being affected by the repeater noise get larger. 
After the fidelity goes under the threshold of EPP, 
the fidelity continues to decrease as the quantum repeater protocol proceeds. 
So, after crossing the threshold, 
$\Delta F_{ES}+\Delta F_{EPP}+\Delta F_{RN} \geq \Delta F_{RN}$, 
where $\Delta F_{ES}, \Delta F_{EPP}$ and $\Delta F_{RN}$ are 
the fidelity decreases due to ES, EPP and the repeater noise, 
respectively, during an iteration in the recursive execution of the quantum repeater protocol. 
The dumping due to the repeater noise is exponential according to the exponential dumping assumption. 
So, the overall dumping is exponential according to the equation above. 
\end{proof}
Of course, there may be many other quantum repeater protocols.
However, in general, our result holds for protocols
as long as the classical communication whose distance is proportional to 
the distance between a sender and a receiver is used in the protocols.

So, it is important to suppress the repeater noise as small as possible.
As the repeater noise gets smaller, 
the range where quantum repeater can improve the bit rate becomes wide.
\section{Security}\label{sec6}
The security proof can be done with simple idea.
Let $I_{E}$ be the amount of eavesdropper's information and 
$P_{cont}$ be the person who controls the quantum repeater unit 
and Eve be an eavesdropper. 
The following relationship holds for $I_{E}$. 
($I_{E}$ in original QKD)$\geq$ 
($I_{E}$ in QKD with quantum repeater, where $P_{cont}$ is Eve)$\geq$ 
($I_{E}$ in QKD with quantum repeater, where $P_{cont}$ is except Eve). 

The reason is as follows. 
For QKD with quantum repeater, Eve can get more information (or equal at least) when 
he controls repeater unit more than when he does not. 
So, ($I_{E}$ in QKD with quantum repeater, where $P_{cont}$ is Eve)$\geq$ 
($I_{E}$ in QKD with quantum repeater, where $P_{cont}$ is except Eve). 
Thus, it is sufficient to prove the security when repeater unit is controlled by Eve. 
Here we deal with QKD protocols where operations for quantum repeater protocol 
can be done within attacks allowed for Eve in original QKD protocol. 
Unconditionally secure protocols, such as BB84\cite{BB84}, belong to this category 
because Eve is allowed to do almost every quantum operations as attacks. 
So, when repeater unit is controlled by Eve, 
operations for quantum repeater can be considered as a part of Eve's 
attacks allowed in original QKD.
Then, ($I_{E}$ in original QKD)$\geq$ 
($I_{E}$ in QKD with quantum repeater, where $P_{cont}$ is Eve).
Thus, we can turn the proof of QKD with quantum repeater into the proof of original QKD. 

\section{Summary}\label{sec7}
We demonstrated that the bit rate of QKD with quantum repeater decreases 
asymptotically exponentially with respect to the communication distance 
when the repeater noise is taken into account. 
This is because EPP can not work when the length of EPR pairs exceed the threshold.
In contrast, quantum repeater protocol works when the length of EPR pairs does not exceed the threshold. 
This threshold depends on the largeness of the repeater noise. 
So, it is important to suppress the repeater noise in order to enlarge the range where 
quantum repeater is effective.
Besides, we showed abstract of proof that QKD with quantum repeater is secure.

\bibliographystyle{jplain}
\bibliography{myrefs,crypto,repeater}

\end{document}